\documentclass[letterpaper]{article}

\usepackage{natbib,alifeconf}  
\usepackage[utf8]{inputenc}
\usepackage{amsfonts}
\usepackage{amsmath}
\usepackage{dsfont}
\usepackage{booktabs}
\usepackage{subcaption}
\usepackage{hyperref}
\urldef\projecturl\url{https://hugocisneros.com/ALIFE-Paper-2020/}

\title{Visualizing computation in large-scale cellular automata}
\author{Hugo Cisneros$^{1}$,  Josef Sivic$^{1,2}$, Tomas Mikolov$^{1}$\\
  \mbox{}\\
  $^1$CIIRC --- Czech Institute of Informatics, Robotics and Cybernetics, Czech
  Technical University in Prague.\\
  $^2$WILLOW project, Département d’Informatique de l’École Normale
  Supérieure \\ ENS/INRIA/CNRS UMR 8548, PSL Research University.} 

\begin{document}
\maketitle

\begin{abstract}
  Emergent processes in complex systems such as cellular automata can perform
  computations of increasing complexity, and could possibly lead to artificial
  evolution. Such a feat would require scaling up current simulation sizes to
  allow for enough computational capacity. Understanding complex computations
  happening in cellular automata and other systems capable of emergence poses
  many challenges, especially in large-scale systems. We propose methods for
  coarse-graining cellular automata based on frequency analysis of cell states,
  clustering and autoencoders. These innovative techniques facilitate the
  discovery of large-scale structure formation and complexity analysis in those
  systems. They emphasize interesting behaviors in elementary cellular automata
  while filtering out background patterns. Moreover, our methods reduce large 2D
  automata to smaller sizes and enable identifying systems that behave
  interestingly at multiple scales.
\end{abstract}

\section{Introduction}
Cellular automata (CA) have been extensively studied since the 1960s. Originally
designed and studied to create artificial evolution from self-replication
\citep{vonneumannTheorySelfreproducingAutomata1966,
  langtonSelfreproductionCellularAutomata1984}, previously studied cellular
automata simulations were often of relatively modest sizes. Only specific rules
with repetitive or predictable dynamics such as John Conway's Game of Life
\citep{gardnerMathematicalGames1970} have been scaled up to larger grid sizes
($10^4 \times 10^4$ or more cells).

For complex phenomena such as artificial evolution to exist and be open-ended
within those simulated worlds, there needs to be sufficient ``capacity'' --- a
large enough state-space. In nature, complex and significantly different
dynamics often arise from uniform laws at a smaller
scale~\citep{andersonMoreDifferent1972}. It seems unlikely that such complex
processes, like artificial evolution, could happen in too small CAs because
higher order dynamics do not have enough capacity to emerge. However, several
issues arise when scaling CAs to large sizes: \vspace{-5pt}
\begin{itemize}
\item Time complexity rapidly becomes a bottleneck. Updating a large number of
  cells is costly. Tricks such as caching of some of the computations can help,
  but do not always improve performance
  significantly~\citep{gosperExploitingRegularitiesLarge1984}.
  \vspace{-6pt}
\item Memory complexity can also become an issue when dealing with numerous
  states, and especially grids in 3 dimensions and more. In that case, even the
  underlying rule of the system cannot be stored within reasonable memory
  capacity.
  \vspace{-6pt}
\item Visual inspection of these large grids is infeasible. Studying CA
  complexity is rendered difficult by the highly variable nature of emergent
  processes. It is especially the case for large-scale systems.
  \vspace{-5pt}
\end{itemize}
When working with such large systems, it is less relevant to focus on the local
behaviors at the single cell level. This is similar to other complex systems
like the weather, in which behaviors of individual atoms in a cloud are
irrelevant to large-scale air mass movements. Much richer behaviors can be
observed from studying large patterns' formation and their evolution. This
should also hold true for CAs; we further discuss this question in
\nameref{sec:conclusion}.
\vspace{-6pt}
\begin{figure}[th]
  \centering
  \includegraphics[width=.93\linewidth]{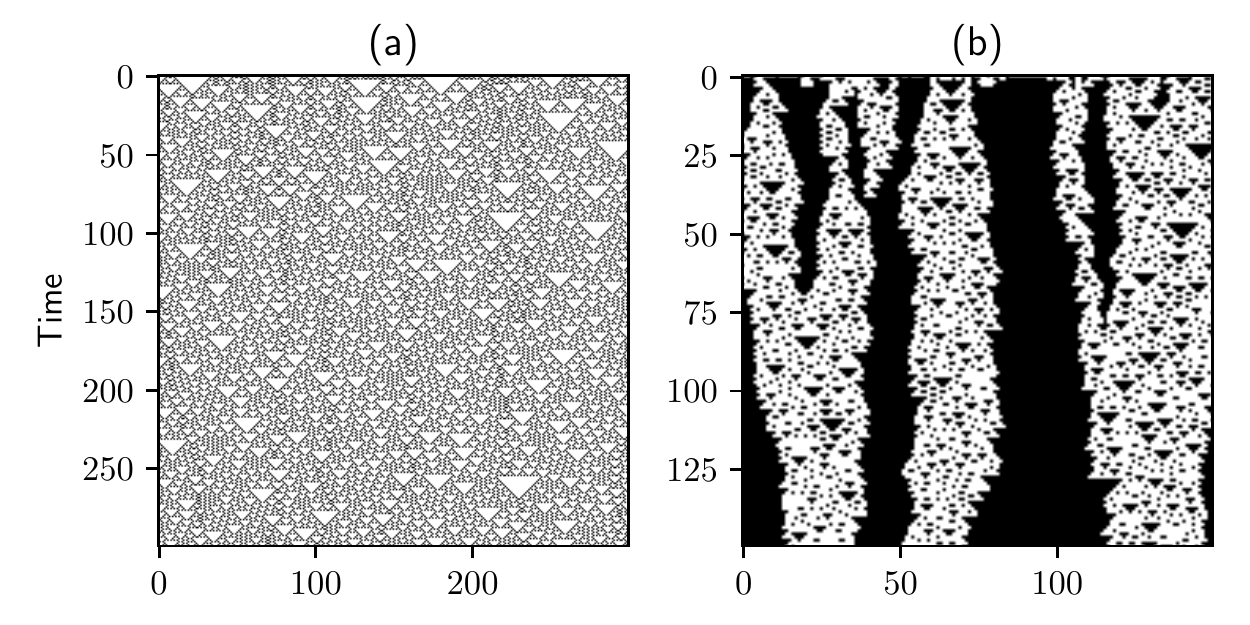}
  \caption{\label{fig:rule18_small} \textbf{Hidden structures in rule 18 are
      uncovered by filtering the space-time diagram with our frequency
      histogram-based method}. \textbf{(a)} shows 300 timesteps of a randomly
    initialized rule 18 simulation. Notice the complex structures made visible
    in \textbf{(b)} with our method.}
\end{figure}
\vspace{-6pt}
In this paper, we investigate techniques which can help us visualize large
space-time diagrams of CAs. We demonstrate that simple clustering and
coarse-graining techniques can be used in order to perceive structures which
cannot emerge on smaller grids. This is also useful for disordered cellular
automata with hidden structures as it is the case for the elementary cellular
automaton rule 18, illustrated in Figure~\ref{fig:rule18_small} --- more details
in~\nameref{sec:results}.

Reducing large grids to smaller sizes while preserving interesting behaviors
such as pattern formation is essential to apply to these CAs complexity metrics
designed to work on modestly sized
grids~\citep{grassbergerQuantitativeTheorySelfgenerated1986,
  zenilCompressionBasedInvestigationDynamical2010,
  soler-toscanoCalculatingKolmogorovComplexity2014,
  zenilTwodimensionalKolmogorovComplexity2015}. Common metrics of complexity are
often limited by the number of components in the systems (number of cells in a
CA grid, timesteps, etc.) or may not be effective when small scale patterns are
less relevant than large-scale ones.

\section{Related work}

Previous work on coarse-graining cellular automata focused either on conserving
the main computational properties of CA rules through exact coarse-graining or
on filtering interesting behaviors without reducing the amount of computations.
Our work both highlights interesting behaviors and compresses the
representation, which we argue are necessary to study complexity in large
cellular automata.

\subsection{Coarse-graining in cellular automata}
Coarse-graining is an approximation procedure used to speed up computations in
systems made of many components. It originated
in~\cite{levittComputerSimulationProtein1975} and is now widely used in physics
to model complex systems at various granularity levels, and is successful at
modeling bio-molecules \citep{potoyanRecentSuccessesCoarsegrained2013,
  ingolfssonPowerCoarseGraining2014, kmiecikCoarseGrainedProteinModels2016}.

Exact coarse-graining of elementary cellular automata (ECA) has been
investigated extensively
in~\cite{israeliComputationalIrreducibilityPredictability2004,
  israeliCoarsegrainingCellularAutomata2006}. Authors found ways of rewriting
one-dimensional CA rules into each other through coarse-graining of the
transition rule. They built a graph of equivalence of all 256 ECA and identified
some rules that do not admit any computational reduction. This indicates that
some cellular automata are accomplishing fundamentally more computations than
others.

\subsection{Filtering}

Filtering cellular automata (CA) was introduced to reduce a CA's behavior to its
most relevant parts. The goal is to extract relevant irregularities from a CA's
space-time diagram. Seminal work
by~\cite{hansonAttractorbasinPortraitCellular1992,
  hansonComputationalMechanicsCellular1997} formalized the notion of domains and
coherent structures in cellular automata. They used a set of regular languages to
represent cellular automata dynamics and extract relevant behaviors such as
discontinuities between regular domains or ``particles''.
Figure~\ref{fig:rule110} shows a filtering example for cellular automaton rule
110 --- in Wolfram's numbering.

A filtering method similar to our proposed frequency-based coarse-graining ---
originally presented as a complexity metric for cellular automata --- is
introduced in~\cite{wuenscheClassifyingCellularAutomata1999}. The author
proposes to progressively filter out cells in cellular automata's space-time
diagrams according to read frequency of the rule table. Cells that originated
from frequent rule table lookups are set to a quiescent or null state. The
choice of threshold has to be decided by a user for each rule. Another notable
difference is the method aims at making visualization of gliders easier without
reducing the size of the grid or making more compact representations.

More recent work by~\cite{shaliziAutomaticFiltersDetection2006} uses the
combination of a modified Lyapunov exponent approach with \emph{statistical
  complexity} \citep{shaliziQuantifyingSelfOrganizationOptimal2004} to underline
complex behaviors. However, the first method requires repeated perturbations and
simulations of the system to study its sensitivity.

\subsection{Scaling-up cellular automata}
Hashlife \citep{gosperExploitingRegularitiesLarge1984} and other Game of
Life-specific optimizations enable simulating a large number of cells for
numerous timesteps. Nonetheless, these algorithms essentially exploit input
redundancy. The regularity in patterns allowing such optimizations might
indicate a lack of novel patterns being generated by the system.

This also means that Game of Life-based simulations are computationally
reducible to a much simpler system, indicating that its computations are
inefficient~\citep{wolframNewKindScience2002}. An optimally complex-behaving
computational model should be impossible to predict except when computing its
actual evolution step by step.

In the following, we used coarse-graining as a method for scaling down CAs in
both time and space in order to make visualization of larger patterns and
complex behaviors easier. The underlying fine-scale computations may be
essential for these larger patterns to appear, hence the necessity to keep them.
However, analogous to many natural processes (swarms, chemistry, cells in an
organism, DNA), interesting behaviors might not be observable at the level of
individual components --- or small groups of components (individuals, single
cells or molecules in the examples above). We view coarse-graining as a way to
reduce a cellular automaton's space-time diagram to its most relevant parts
while keeping primary dynamics in the background. The resulting diagram would
ideally be an irreducible system.

\section{Proposed coarse-graining of cellular automata}
For reasons stated above, we introduce coarse-graining methods for cellular
automata that are not reversible --- information is discarded in the process.
This process does not attempt to find shortcuts for the computations of a
cellular automaton, but rather to selects relevant parts of the space-time
diagram and discards information irrelevant to the core behavior. For example, a
standard glider in Game of Life spanning $3\times 3$ cells could be replaced
with a single cell moving diagonally when coarse-graining by a factor 3. This is
because the actual oscillator's dynamics might not be relevant at this coarser
scale.

Coarse-graining is akin to constructing \emph{supercells} from blocks of individual
cells. These supercells are assigned a new state and form a coarser partitioning
of the initial grid which can be studied as its own system. In particular,
complexity metrics or further coarse-graining can be applied to this new grid.

\subsection{Frequency histogram coarse-graining}\label{sec:simple-hier-coarse}

A simple coarse-graining is achieved by mapping blocks to a single
\emph{supercell} state according to the probability of this configuration
appearing, given a previously constructed model. The easiest way to think of it
is with a simple frequency counting model of the distribution of $2\times 2$
blocks in a 2D CA\@. For a 2-state automaton, there are 16 possible supercell
configurations. The simplest model for the occurrence of these blocks is their
empirical frequency. Let us consider a CA with $N$ blocks of $2\times 2$ cells,
let $S^{(in)} = \{\mathtt{0000}, \mathtt{0001}, \mathtt{0010}, \ldots,
\mathtt{1111}\}$ be the set of $2\times 2$ blocks and $s_i \in S^{(in)}$ be a
given supercell. The probability $p_i$ of observing supercell $i$ on a grid $G$
is estimated with
\begin{equation}
  p_i = \dfrac{\text{count}_G(s_i)}{\sum_{j\in S^{(in)}}\text{count}_G(s_j)}
  \label{eq:stat_est}
\end{equation}
where $\text{count}_G(s_i)$ is the number of blocks matching $(s_i)$ in $G$.

Supercells can then be assigned a particular state. We call the corresponding
mapping $f: S^{(in)} \mapsto S^{(out)}$. $S^{(out)}$ can be chosen depending on
the desired output or use. For instance, with $S^{(out)} = \{0, 1\}$ we can
define $f$ to map each supercell $s_i$ as follows:
\begin{align}
  f(i) = \begin{cases}
    \mathtt{0} &\quad\text{if }\ p_i\geq \alpha\\
    \mathtt{1} &\quad\text{if }\ p_i< \alpha
  \end{cases}
        \label{eq:alpha}
\end{align}
where $\alpha$ is a chosen threshold.

\subsubsection{Partitioning the histogram.}
This method can be understood as partitioning the histogram of supercell
frequency. In equation~\eqref{eq:alpha}, supercells with low probability
--- with higher self-information --- are mapped to state \texttt{1} whereas
commonly occurring states are mapped to \texttt{0}.

Choosing a partition of the histogram is equivalent to selecting a suitable
$\alpha$ --- scalar for two output states, or vector $\mathbf{\alpha} =
(\alpha_1, \ldots, \alpha_n)$ for $n$ output states. Therefore, one can map
supercells to any number of target states (three or more) by partitioning the
frequency histogram into any number of bins. Supercell distribution can be
anything between uniform and very unbalanced, with a few supercells being
overwhelmingly represented (background) and only a few occurrences of other
configurations. The chosen partitioning has to deal with both situations equally
well. In the following, we use a uniform partitioning of the area under the
negative log-histogram for elementary cellular automata --- supercells are
divided into two bins of equal summed negative logarithmic probability. For 2D
CAs, we use the same method but with quadratic partition of the histogram
($1/k^2$ instead of $1/k$, with $k$ the number of output states, chosen because
of better visual results).

\subsubsection{Dithering.}\label{sec:dithering}
\begin{figure}[th]
  \centering
  \begin{subfigure}{.32\linewidth}
    \fbox{\centering
    \includegraphics[width=\linewidth]{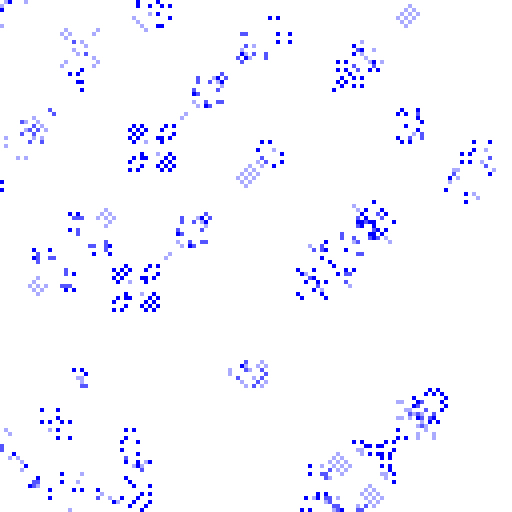}}
    \caption{\label{subfig:normal}Original CA}
  \end{subfigure}
  \hfill
  \begin{subfigure}{.32\linewidth}
    \fbox{\centering
    \includegraphics[width=\linewidth]{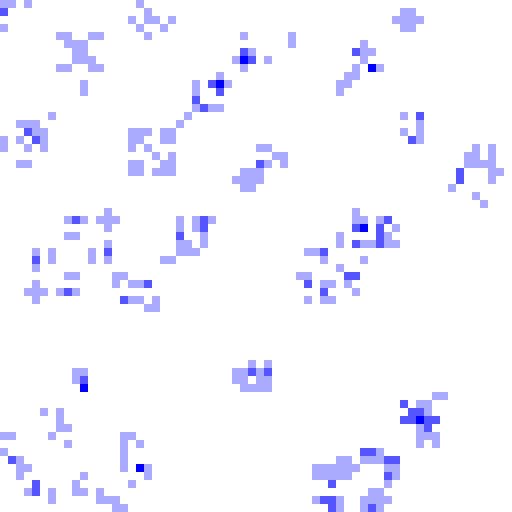}}
    \caption{\label{subfig:no-dithering}w/out dithering}
  \end{subfigure}
  \hfill
  \begin{subfigure}{.32\linewidth}
    \fbox{\centering
    \includegraphics[width=\linewidth]{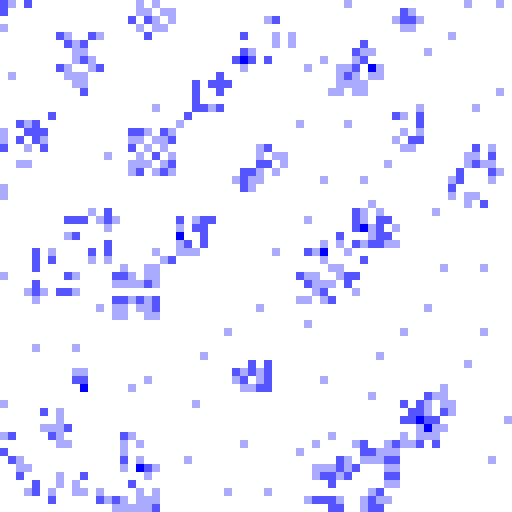}}
    \caption{\label{subfig:dithering}with dithering}
  \end{subfigure}

  \caption{\label{fig:close-up}\textbf{Close-up view of coarse-graining effects
      on a 4-states CA rule} (1 shade of blue per state). Both coarse-graining
    methods conserve many of the interesting structures. Dithering introduces
    additional artifacts on regular backgrounds. Fig.~\ref{subfig:normal} shows
    actual states in the CA simulation on a $128 \times 128$ grid.
    Fig.~\ref{subfig:no-dithering} is a coarse-grained version
    of~\ref{subfig:normal} with histogram coarse-graining, the grid is $64
    \times 64$ cells.\ref{subfig:dithering} is obtained with histogram
    coarse-graining and dithering (see \nameref{sec:dithering}).}
\end{figure}

Histogram partitioning introduces another set of parameters to be manually
tuned, adding complexity to the procedure. An alternative way to produce an
output image from the histogram is to use dithering. Dithering is an image
processing technique commonly used to reduce large visual artifacts induced by
quantization errors. Noise is added to the image during the quantization process
to make the average local value of a set of pixels as close to their target
continuous value as possible. The resulting image is created so as to match
target continuous values with discrete values only --- cell states in the grid.
It can be seen as another way of partitioning the histogram with variable
thresholds that depend on a running quantization error.
Figure~\ref{fig:close-up} shows a comparison of dithering and regular histogram
partitioning (Floyd–Steinberg's algorithm was used
\citep{floydAdaptiveAlgorithmSpatial1976}).

\setlength{\fboxsep}{0pt}

\subsubsection{Visualization.}
\begin{figure}[th]
  \centering
  \begin{subfigure}{.48\linewidth}
    \fbox{
    \centering
    \includegraphics[width=\linewidth]{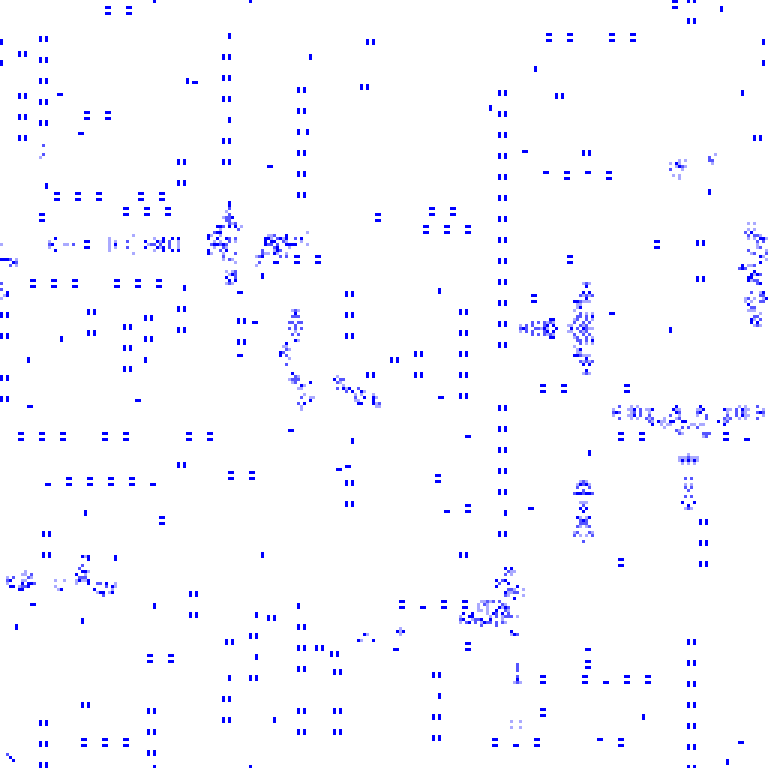}}
    \caption{\label{subfig:gol}Base grid}
  \end{subfigure}
  \hfill
  \begin{subfigure}{.48\linewidth}
    \fbox{
    \centering
    \includegraphics[width=\linewidth]{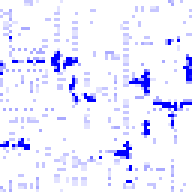}}
    \caption{\label{subfig:gol_cg}Coarse-grained}
  \end{subfigure}

  \caption{\label{fig:gol_comparison}\textbf{Side-to-side comparison of a CA
      simulation and its coarse-grained version}. The first simulation is $256
    \times 256$ cells and the second has been coarse-grained to $64\times 64$.
    Notice the interesting patterns on Figure~\ref{subfig:gol} are hardly
    distinguishable. They are highlighted by histogram-based coarse-graining in
    Figure~\ref{subfig:gol_cg}.}
\end{figure}

One advantage of this frequency histogram-based method is that it naturally
highlights rarer events in the simulation grid, creating a ``heatmap'' of the
simulation's activity. Since we sort supercells according to their observed
frequency, the right choice of colors --- e.g.\ progressively darker gradient
--- can lead to automatic highlighting of active regions of a cellular
automaton. Figure~\ref{fig:gol_comparison} shows the same simulation both
unprocessed and downscaled by a factor of 4 with coarse-graining. Although much
coarser, Figure~\ref{subfig:gol_cg} is more readable than the base version,
which is helpful when dealing with large grids\footnote{Several figures in this
  paper have animated versions, accessible at the paper's project page
  \projecturl}.

\subsubsection{Hierarchical coarse-graining.}
The above procedure can be applied recursively to the same cellular automaton or
with larger block sizes to get a progressively coarser representation. Since
information is systematically discarded in the process, it cannot be applied any
number of times. For this reason, many 2D CAs exhibiting interesting behaviors at
the micro-level but not at the macro-level have no remaining visible structure
after reducing their scale several times with this method.

Because a simple model like frequency counting can be estimated quickly,
hierarchical coarse-graining is easily applied to large grids, reducing the size
by a factor of $n$ (block size) every time. For instance, this property makes it
suitable to search the cellular automata rule space for CAs behaving
interestingly at multiple coarse-graining levels simultaneously.

\subsection{Clustering}

Another way to convert blocks of cells for coarse-graining is to distribute
these blocks into a small number of clusters, where each group becomes the new
coarse state.

Several distance functions may apply here, the most natural of which being
Hamming distance, which measures how many states differ between two
positions~\citep{hammingErrorDetectingError1950}. It is defined for two strings
of equal length $n$, $s_1 = [s_{(1, 1)}, \ldots, s_{(1,n)}]$ and $s_2 =
[s_{(2,1)}, \ldots, s_{(2,n)}]$, as the number of positions where the two
strings differ:
\begin{align}
  \sum_{k = 1}^n \mathds{1}\left\{\ s_1^{(k)}  \neq s_2^{(k)}\right\}.
\end{align}

A supercell of $N\times N$ cells of a CA can be converted into a string to be
compared to other blocks with the Hamming distance. For CAs, we limit ourselves
to strings of digits representing states, i.e. $s_{(i,j)} \in \mathbb{N}$. We
use a vanilla implementation of the K-means algorithm where clusters' centers
are computed using a continuous average of position vectors rounded to nearest
integer values. Clusters are initialized with randomly selected observations.

\subsection{Autoencoders for coarse-graining}

Instead of just relying on the amount of information of a given supercell's
configuration, one can also try to automatically find a relevant representation
with dimensionality reduction methods. Autoencoders are neural networks composed
of an encoder part and a decoder part, originally designed to identify principal
components of a collection of data
points~\citep{baldiNeuralNetworksPrincipal1989,
  hintonConnectionistLearningProcedures1989,
  kramerNonlinearPrincipalComponent1991}. An encoder neural network converts
data to a \emph{latent} vector of smaller dimension than the original input.
Then, a decoder neural network reconstructs a vector with the same dimension as
the input from this encoded \emph{latent} representation.These models can
automatically find an optimal constrained representation through minimizing a
reconstruction loss between the original input and the reconstructed output.

We denote the encoder network with $E$ and the decoder network with $D$. We frame
the reconstruction problem as a $N$ class classification problem with multiple
components --- one class per input state, one component for each cell of the $K$
cells in a block. The reconstruction loss is the component-wise cross-entropy
between the state of each input cell and the reconstructed state after $D \circ
E$ is applied.

\begin{figure}[th]
  \centering
  \includegraphics[width=\linewidth]{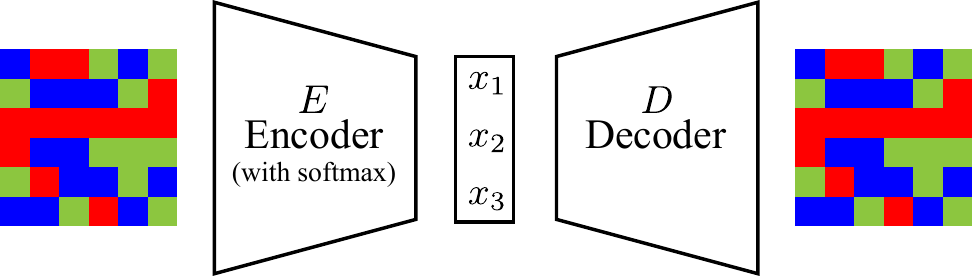}
  \caption{\label{fig:autoencoder} \textbf{Diagram of the autoencoder
      architecture used for coarse-graining}. A block of $6\times 6$ cells is
    encoded in a vector of fixed dimension. There are 3 components in the
    example. They can either represent a RGB color or a 3 states smaller
    automaton.}
\end{figure}

Figure~\ref{fig:autoencoder} illustrates the autoencoder layout for
coarse-graining. By adjusting the block size and dimension of the encoded
vector, one can influence the amount of information conserved during encoding.
Naturally, smaller blocks will be more easily represented in lower dimension.

The encoder has a softmax layer to ensure the coded state's components sum to
one. Therefore, one can view this coded supercell as a mixture of states which
can either be kept as is or converted to a discrete state by keeping the maximal
component only. They are trained with stochastic gradient descent until
convergence.

\section{\label{sec:results}Results}

We evaluate our proposed coarse-graining methods in the following two different ways:
\begin{itemize}
\item We compare our results on elementary cellular automata (ECA) to previous
  works on particle and domain filtering.
\item We use a metric which evaluates complexity of CAs introduced
  in~\cite{cisnerosEvolvingStructuresComplex2019} in order to compare our
  methods' complexity metric scores of the coarse-grained systems and contrast
  the scores against a standard image processing baseline that computes local
  average of neighbouring cells followed by downscaling the grid. Using the
  complexity metric we measure to what extent the interesting behavior of
  cellular automata is conserved after coarse-graining compared to this image
  processing baseline.
\end{itemize}

In the following we begin by showing that a simple histogram-based
coarse-graining is effective at detecting structures (such as gliders) in ECA
space-time diagrams. Our method achieves results comparable with previous work,
while being simpler to apply.

\subsection{Domains and filtering}

In the space-time diagrams of cellular automata, moving structures such as
gliders are embedded in uniform or periodic backgrounds, or ``domains''. This
domain is different depending on the rule: some ECAs have uniform backgrounds,
checkerboard backgrounds or more complicated patterns (e.g.\ rule
110).~\cite{crutchfieldTurbulentPatternBases1993} also identified chaotic
domains, which cannot support regular gliders but have ``walls'' and
``particles''. Those correspond, respectively, to boundaries between two chaotic
domains and propagating defects (localized structures with a pattern different
from the domain) within a domain.

Our proposed coarse-graining methods offer interesting perspectives to filter
cellular automata's space-time diagrams, which enables identifying gliders and
studying the formation of large-scale patterns. We find that a simple histogram
coarse-graining achieves results comparable to those reported
in~\cite{hansonAttractorbasinPortraitCellular1992,
  elorantaKinkCellularAutomaton1992, hansonComputationalMechanicsCellular1997,
  wuenscheExploringDiscreteDynamics2011} for ECA rules 18 and 54. A similar
approach was undertaken in \citep{wuenscheClassifyingCellularAutomata1999} in
which the authors used the entropy of rule table lookup frequencies to filter
out regular domains in the space-time diagrams of cellular automata and to
identify gliders and domain boundaries. However, Wuensche's approach described
in~\cite{wuenscheClassifyingCellularAutomata1999} does not attempt to downscale
space-time diagrams.

\subsection{Results on elementary cellular automata}

\begin{figure}[th]
  \centering
  \includegraphics[width=\linewidth]{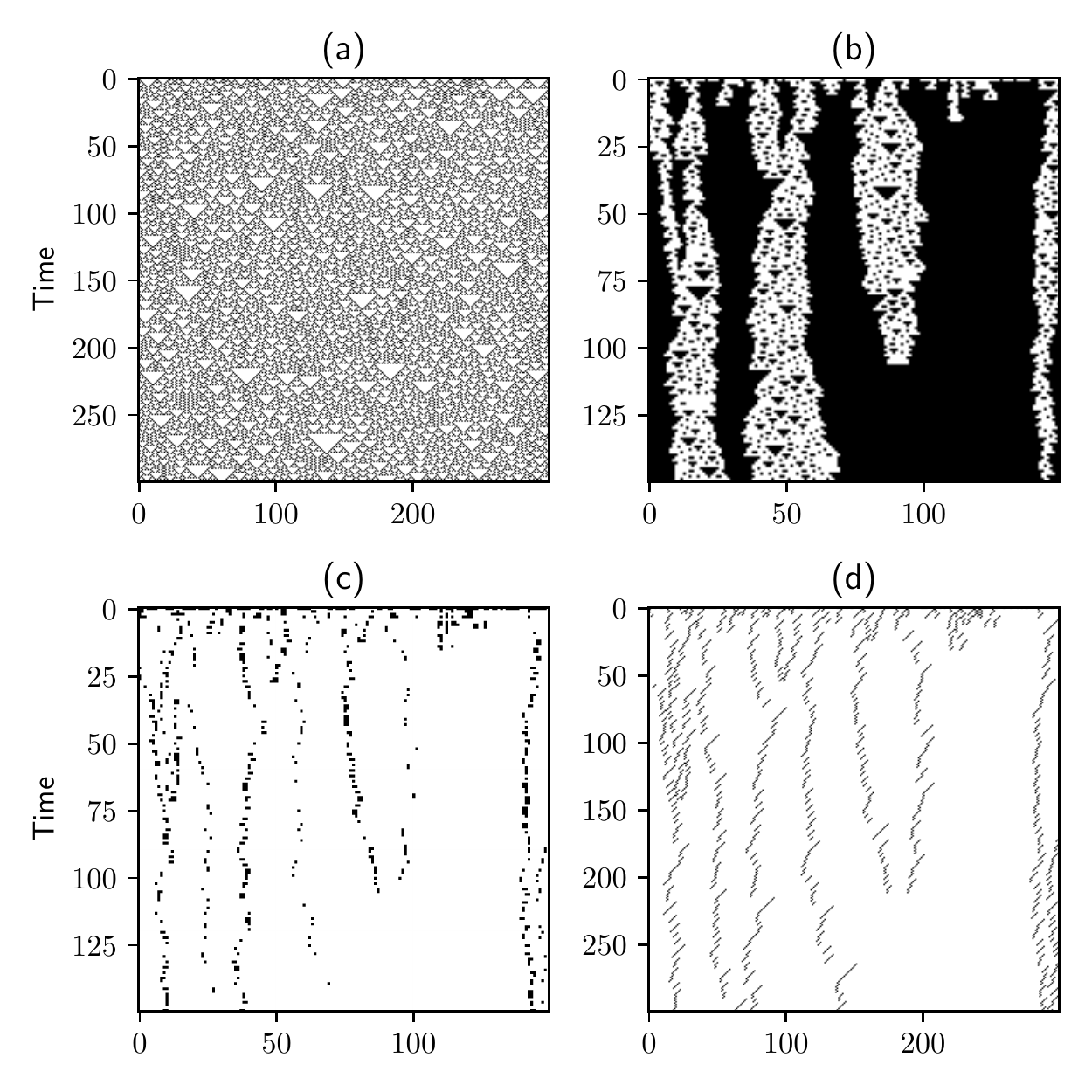}
  \caption{\label{fig:rule18} \textbf{Space-time diagrams for rule 18 in
      elementary cellular automata.} \textbf{(a)} Standard rule 18 space-time
    diagram, starting from a random position. \textbf{(b)} Filtered domain with
    our frequency coarse-graining (even). \textbf{(c)} Our domain boundaries
    extracted from the filtered domains in \textbf{(b)}. \textbf{(d)} Domain
    boundaries computed according to
    \citep{hansonAttractorbasinPortraitCellular1992}. Note that \textbf{(a)}
    shows semi-chaotic behaviour, which is hard to interpret, whereas our method
    \textbf{(b)} highlights distinct domains within the disordered space-diagram
    in (a). The detected domains and domain boundaries from previous work (d)
    and ours (c) are very similar.}
\end{figure}

\begin{figure}[th]
  \centering
  \includegraphics[width=\linewidth]{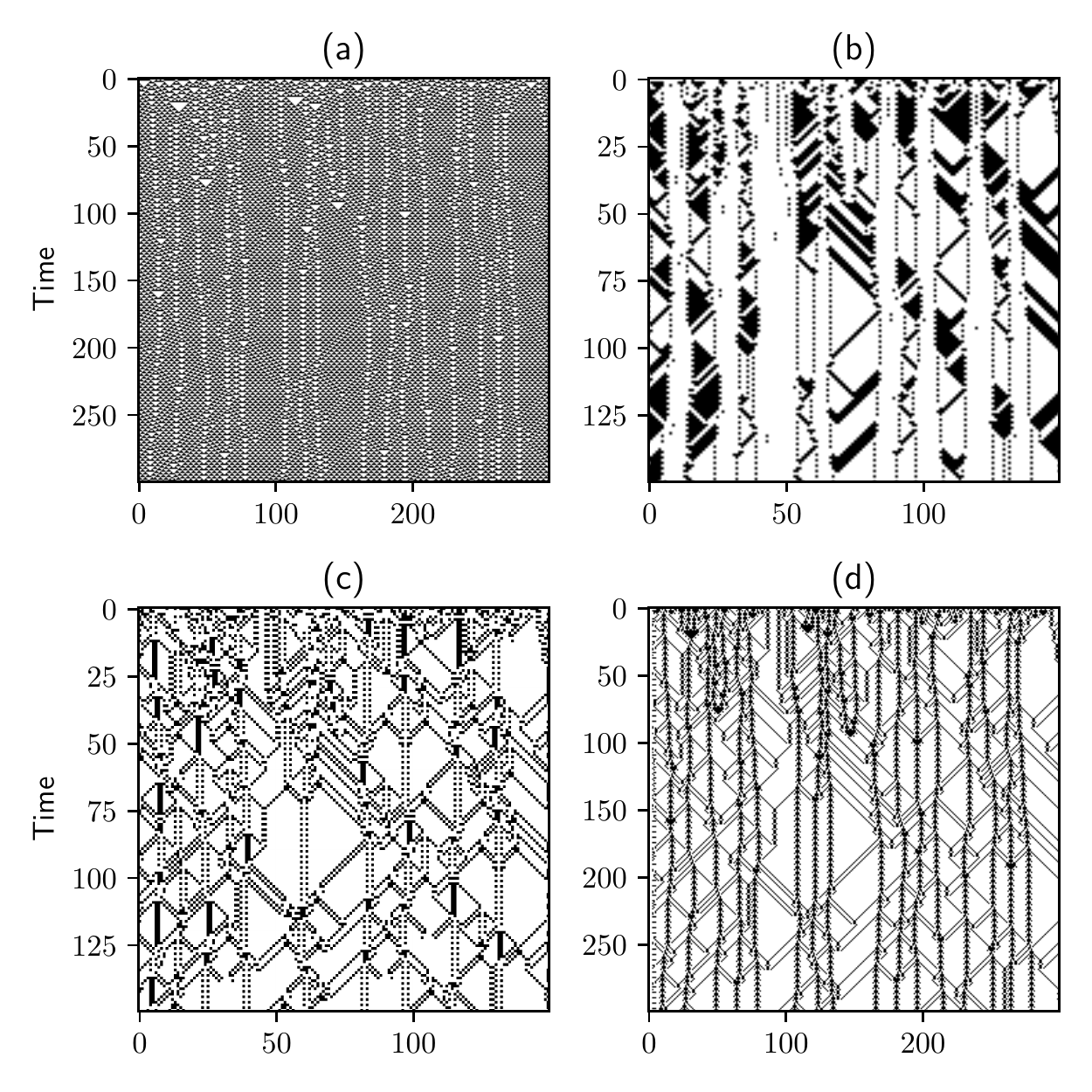}
  \caption{\label{fig:rule54} \textbf{Space-time diagrams for rule 54}.
    \textbf{(a)} Space-time diagram of standard rule 54, starting from a random
    position. \textbf{(b)} Filtered domain with our frequency coarse-graining
    (even). \textbf{(c)} Particles filtered from the domains in \textbf{(b)
      using our method}. \textbf{(d)} Domain boundaries computed using
    computational mechanics \citep{hansonComputationalMechanicsCellular1997}.
    Please note that particles are detected equally well using computational
    mechanics (d) and our (simpler) frequency-based method (c). Some close-by
    particle trails are merged using our method.}
\end{figure}

We apply frequency histogram-based coarse-graining on elementary cellular
automata (ECA) and obtain space-time diagrams with suppressed background
domains. Resulting partitions of ECAs' space-time diagram are similar to results
reported by~\cite{hansonAttractorbasinPortraitCellular1992,
  hansonComputationalMechanicsCellular1997}. Figure~\ref{fig:rule18}\textbf{(a)}
and~\ref{fig:rule54}\textbf{(a)} show the space-time diagrams of rules 18 and 54
with random initialization. Boundaries between different background patterns in
both Figures were obtained with coarse-graining; they are similar to boundaries
obtained by Hanson and Crutchfield. We also observe the propagation of many of
the same particles and defects without any prior information about the cellular
automaton rule.

Particles in Rule 54 have been used to implement
computations~\citep{boccaraParticlelikeStructuresTheir1991,
  pivatoSpectralDomainBoundaries2007,
  martinezCompleteCharacterizationStructure2014}. Because the presented
reduction reduces the size of the grid, it can merge some of those particles,
sometimes resulting in ambiguities and gaps. However, our goal here is not to
precisely describe particle interactions in order to manipulate or construct
complex computations manually. Underlying computations described in the works
above are still happening within our reduced CA simulation. We consider the
\emph{apparent} destruction of some of these fine-scale details acceptable in
order to discover larger-scale complex behavior.

Our method is also arguably much simpler than computational mechanics (used by
Hanson and Crutchfield) which requires some reverse-engineering of the rule and
the construction of a finite-state transducer to generate output symbols.
Although full automation has been demonstrated, this method introduces
significant overhead~\citep{rupeLocalCausalStates2018}. On the other hand, our
method is sensitive to the quality of statistical estimation of the frequency
histogram (see equation~\eqref{eq:stat_est}) and needs enough input examples to
achieve a reasonable result --- examples in Figure~\ref{fig:rule18}
and~\ref{fig:rule54} used simulations with the width of 3000 cells, ran for 6000
timesteps to obtain reliable pattern frequency estimates.

In the Figures, we used the coarse-graining method introduced
in~\nameref{sec:simple-hier-coarse}. Space-time diagrams are coarse-grained by a
factor 2 to a binary automaton --- each cell corresponds to a 2-cell block.
These binary coarse-graining results in the Figures are labeled \textbf{(c)}.
Because of the statistical nature of the domains of these 1D ECA's and the use of
blocks of size 2, filtered domains differ depending on the starting position of 
coarse-graining. We distinguish an odd and even filtered domain.

\begin{figure}[t]
  \centering
  \includegraphics[width=\linewidth]{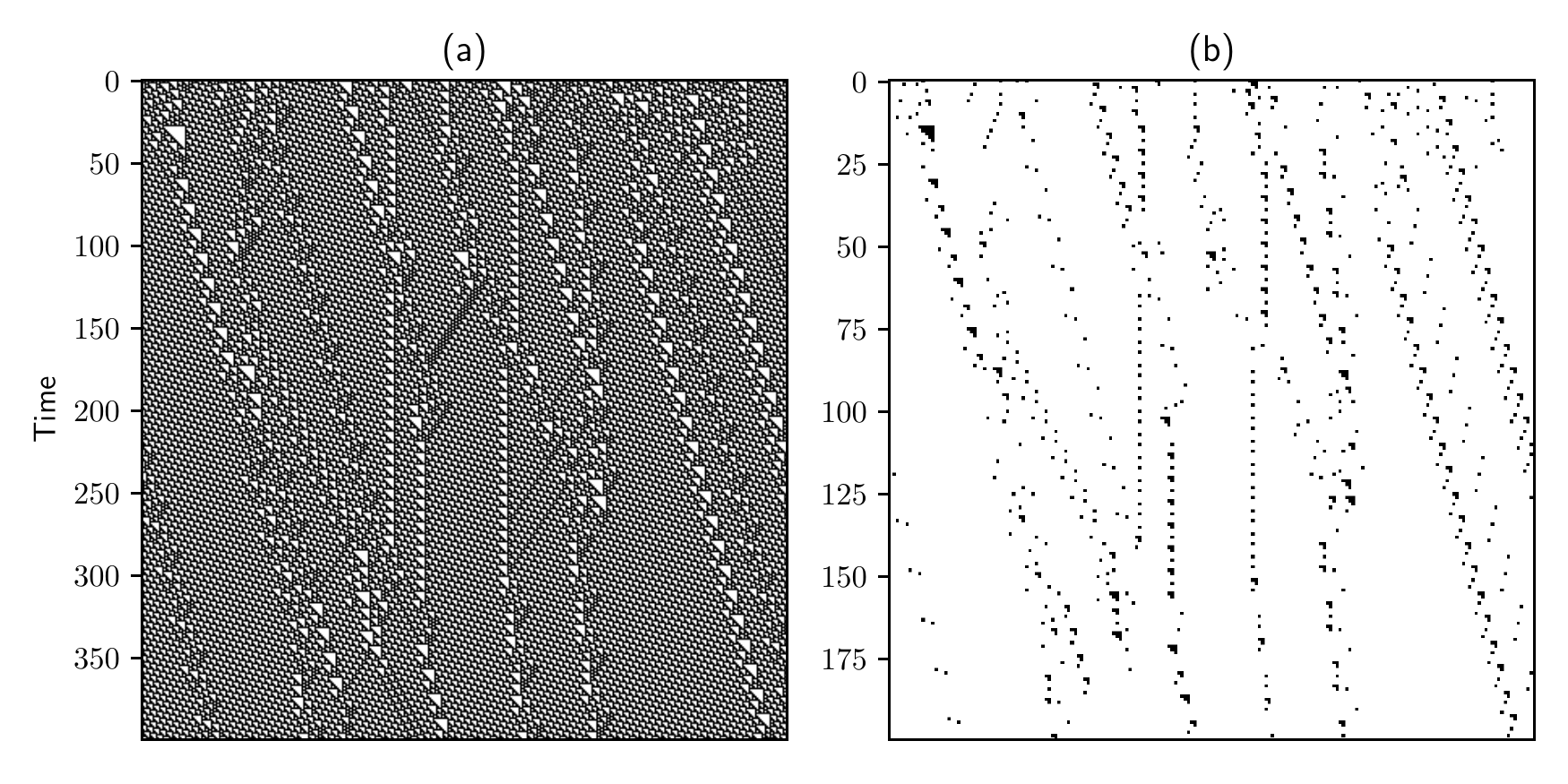}
  \caption{\label{fig:rule110} \textbf{Spate-time diagram of rule 110 (\textbf{a})
      and filtered particles using our histogram-based coarse-graining
      (\textbf{b})}. Structures propagating in time (vertical axis) and space
    (horizontal axis) become clearly visible in \textbf{(b)} as vertical and
    diagonal lines. 
    }
\end{figure}

Figure~\ref{fig:rule18}\textbf{(c)} is obtained by applying the element-wise
\texttt{OR} operator to both the even and odd domain diagrams to merge them into
a single space-time diagram. Figure~\ref{fig:rule54}\textbf{(c)} is obtained by
computing differences between neighboring cells after the filtering process to
highlight lines. Figure~\ref{fig:rule110} is another example showing filtering
of particles in rule 110.

\subsection{Complexity metrics and coarse-graining}

\setlength{\fboxsep}{0pt}
\begin{figure}[th]
  \centering
  \begin{subfigure}{.48\linewidth}
    \fbox{\centering
      \includegraphics[width=\linewidth]{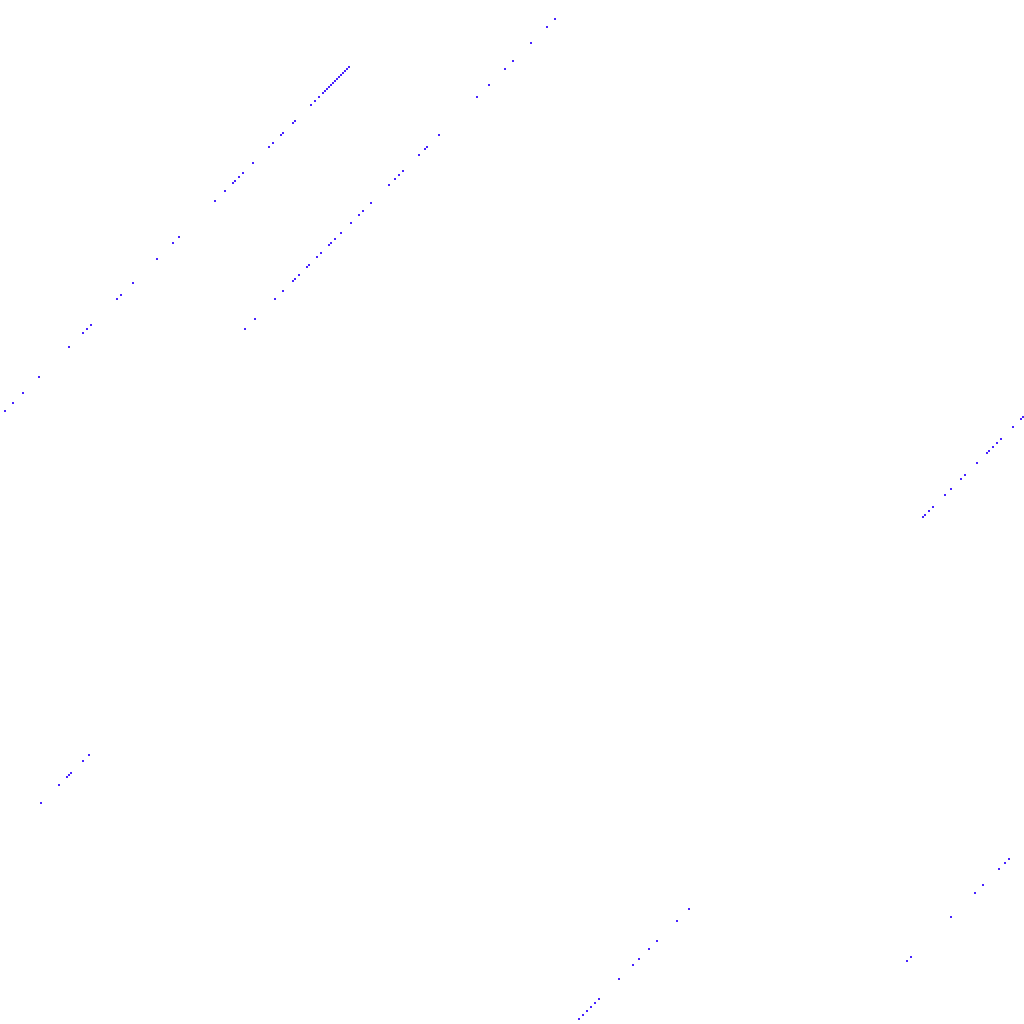}}
    \caption{\label{subfig:downscale3} Downscaling by averaging}
  \end{subfigure}
  \begin{subfigure}{.48\linewidth}
    \fbox{\centering
      \includegraphics[width=\linewidth]{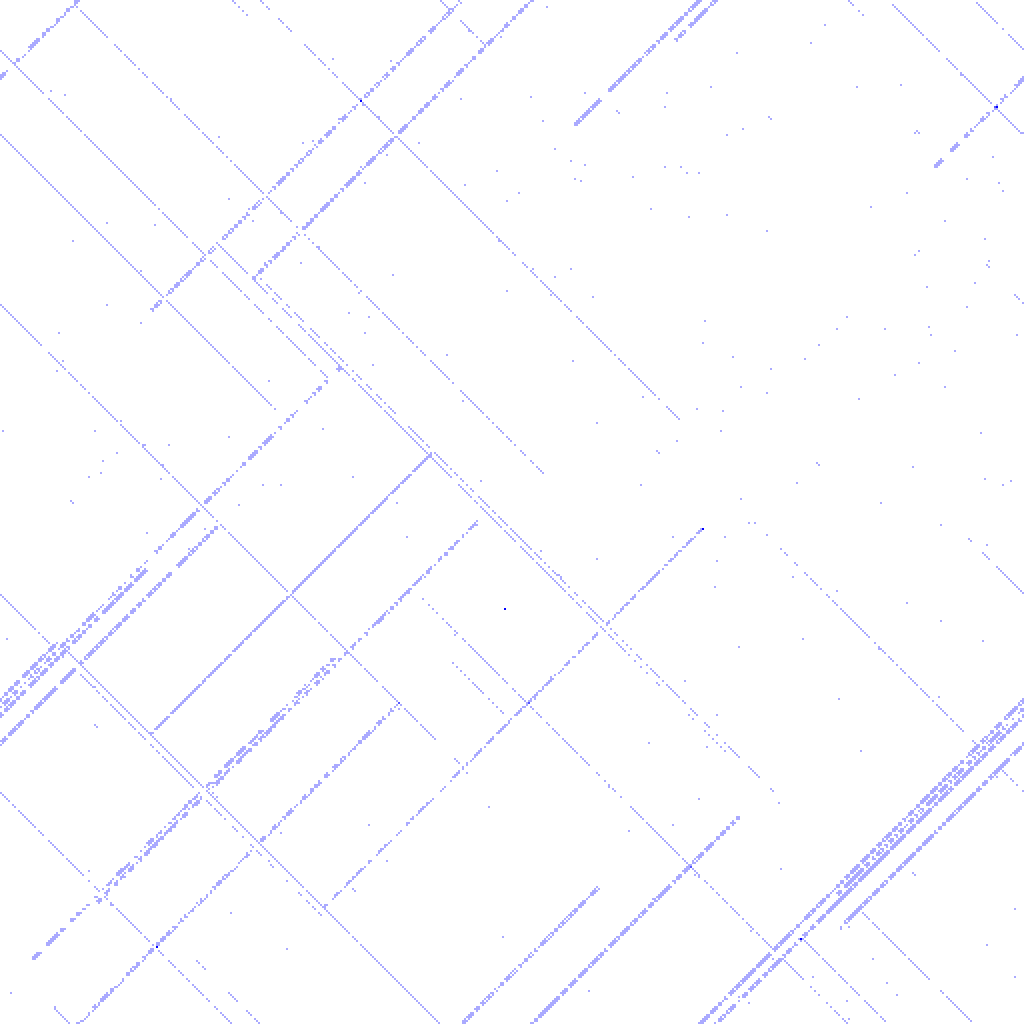}}
    \caption{\label{subfig:histogram3} Histogram}
  \end{subfigure}
  \begin{subfigure}{.48\linewidth}
    \fbox{\centering
      \includegraphics[width=\linewidth]{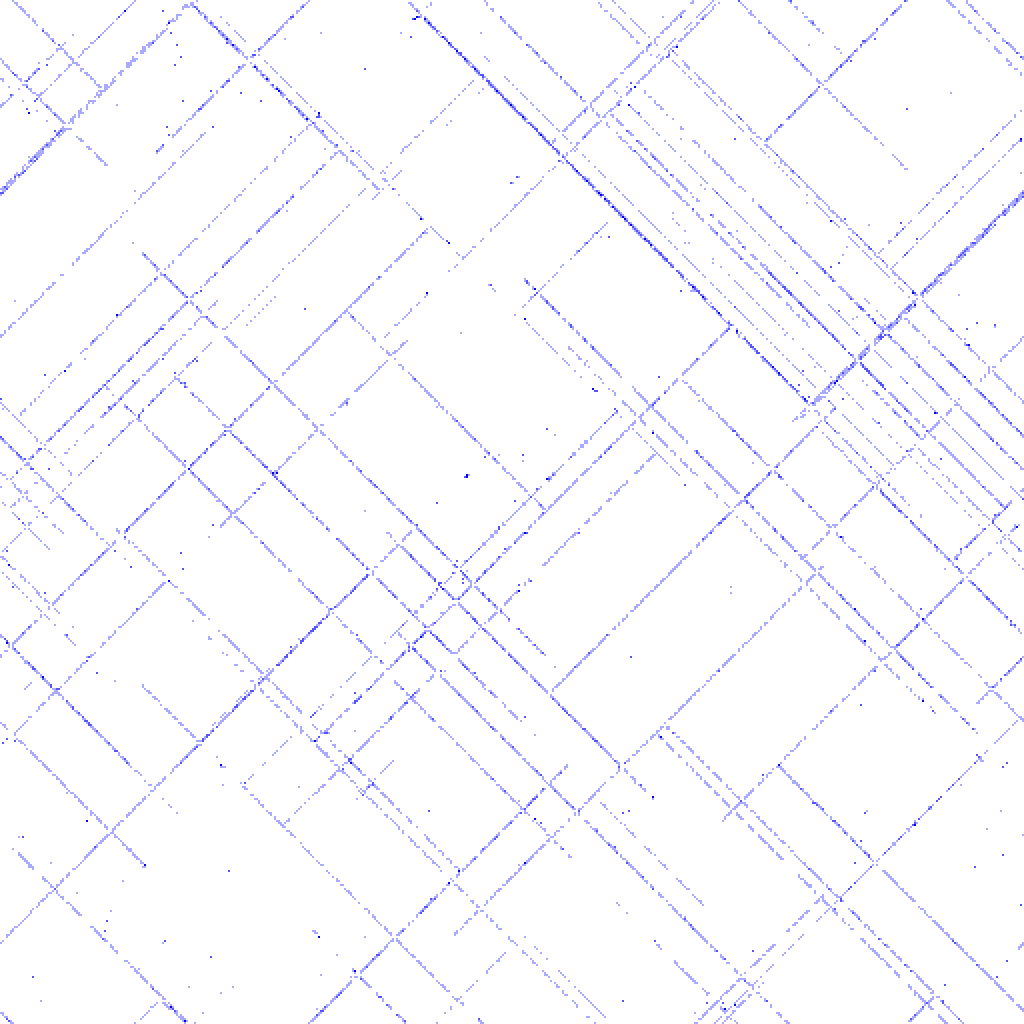}}
    \caption{\label{subfig:kmeans3} K-means}
  \end{subfigure}
  \begin{subfigure}{.48\linewidth}
    \fbox{\centering
    \includegraphics[width=\linewidth]
    {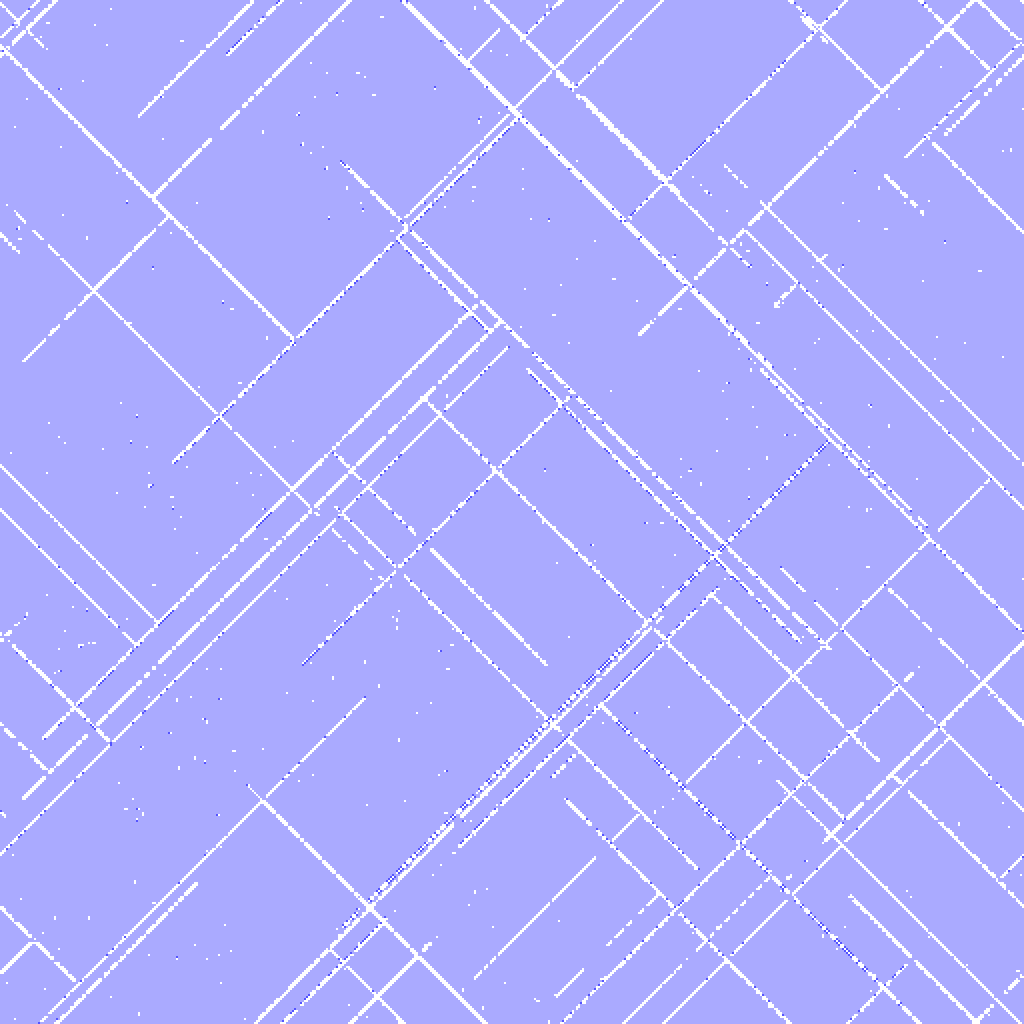}}
    \caption{\label{subfig:autoencoder3}Autoencoder}
  \end{subfigure}
  \caption{\label{fig:qualitative}\textbf{Qualitative comparison of
      coarse-graining methods}. Simulations are on grids of $4096 \times 4096$
    cells coarse-grained to $512 \times 512$. Lines are barely visible with
    downscaling~\ref{subfig:downscale3}, but are visible
    in~\ref{subfig:histogram3}-\ref{subfig:autoencoder3}. Coarse-graining helps
    visualize linear structures that would be hard to see otherwise.}
\end{figure}

Coarse-graining is not only useful for detecting gliders and domains in
space-time diagrams, but also as a tool to visualize large CAs. To evaluate the
quality of our proposed coarse-graining methods, we compare complexity scores
computed according to~\cite{cisnerosEvolvingStructuresComplex2019} for different
coarse-graining methods. This metric was shown to correlate well with a user
study for interesting automata. It uses neural networks to estimate how easy it
is to learn a compressed representation of a CA\@. We also compute the scores on
downscaled CAs as a baseline. Local averaging is used for downscaling, with each
block of $N$ cells being replaced by their average value rounded to the nearest
integer state.

Experiments begin by sampling 3600 cellular automata rules with 3 or 4 states.
We apply the complexity metric on a randomly initialized simulation on a $512
\times 512$ grid of cells. The top 100 rules with the highest complexity scores,
which should correspond to rules with interesting behaviors, are then used for
coarse-graining. We apply coarse-graining on grids of $4096 \times 4096$ cells,
scaling the grid down by a factor of $8$, and compute the complexity metric also
on the reduced grid. Figures reported in Table~\ref{experiments_table} are
percentages of rules still considered interesting (above the selection threshold
for the first step of the process) after coarse-graining. The higher this number
is, the more a method is able to conserve complex and interesting behaviors
after the reduction.

\begin{table}[t!]
  \centering
  \begin{tabular}{ccccc}
    \toprule
       Local-averaging & K-Means & Histogram & Autoencoder\\
     baseline &  & & \\
    \midrule
      19.3\%  & 40.4\% & 82.4\% & 84.2\%\\
    \bottomrule
  \end{tabular}
  \caption{\label{experiments_table}\textbf{Experimental results --- Percentage
      of rules classified as interesting after reduction with our 3 proposed
      methods (K-means, Histogram, Autoencoder), compared to a local averaging baseline.}}
\end{table}

Results in Table~\ref{experiments_table} suggest that using our proposed methods
seems largely beneficial for studying complexity in large systems.
Histogram and autoencoder methods are superior to downscaling using k-means and
local averaging. This could be attributed to the fact that contrary to the
latter two, the histogram and autoencoder both represent well anomalies (rare
events). This is because rare events are explicitly captured and kept by the
histogram method. They also represent useful information that may be kept for
the reconstruction using the autoencoder.

\subsection{Discussion}\label{sec:discussion}

Downscaling by local averaging is not an effective solution to the
coarse-graining problem for several reasons. In particular, it tends to favor
the majority state in a supercell because of the averaging effect. Thin
structures spanning only few cells placed on a uniform background are likely to
disappear after coarse-graining although they may still be relevant with respect
to the large-scale patterns. The histogram-based method explicitly encodes those
more rare events in a supercell, even if their size is relatively small compared
to the supercell size.

\begin{figure}[ht]
  \centering
  \begin{subfigure}{.052\linewidth}
    \fbox{\centering
    \includegraphics[width=\linewidth]{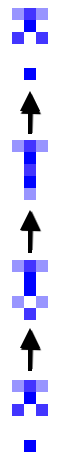}}
    \caption{\label{subfig:single}}
  \end{subfigure}
  \begin{subfigure}{.45\linewidth}
    \fbox{\centering
    \includegraphics[width=\linewidth]{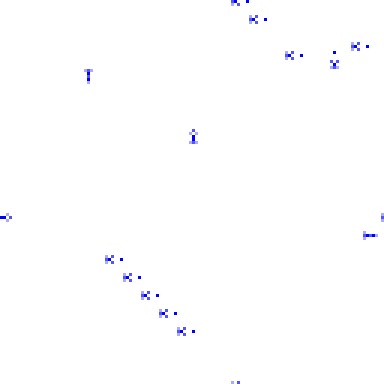}}
    \caption{\label{subfig:mult_glider} $128 \times 128$ cells}
  \end{subfigure}
  \begin{subfigure}{.45\linewidth}
    \fbox{\centering
    \includegraphics[width=\linewidth]{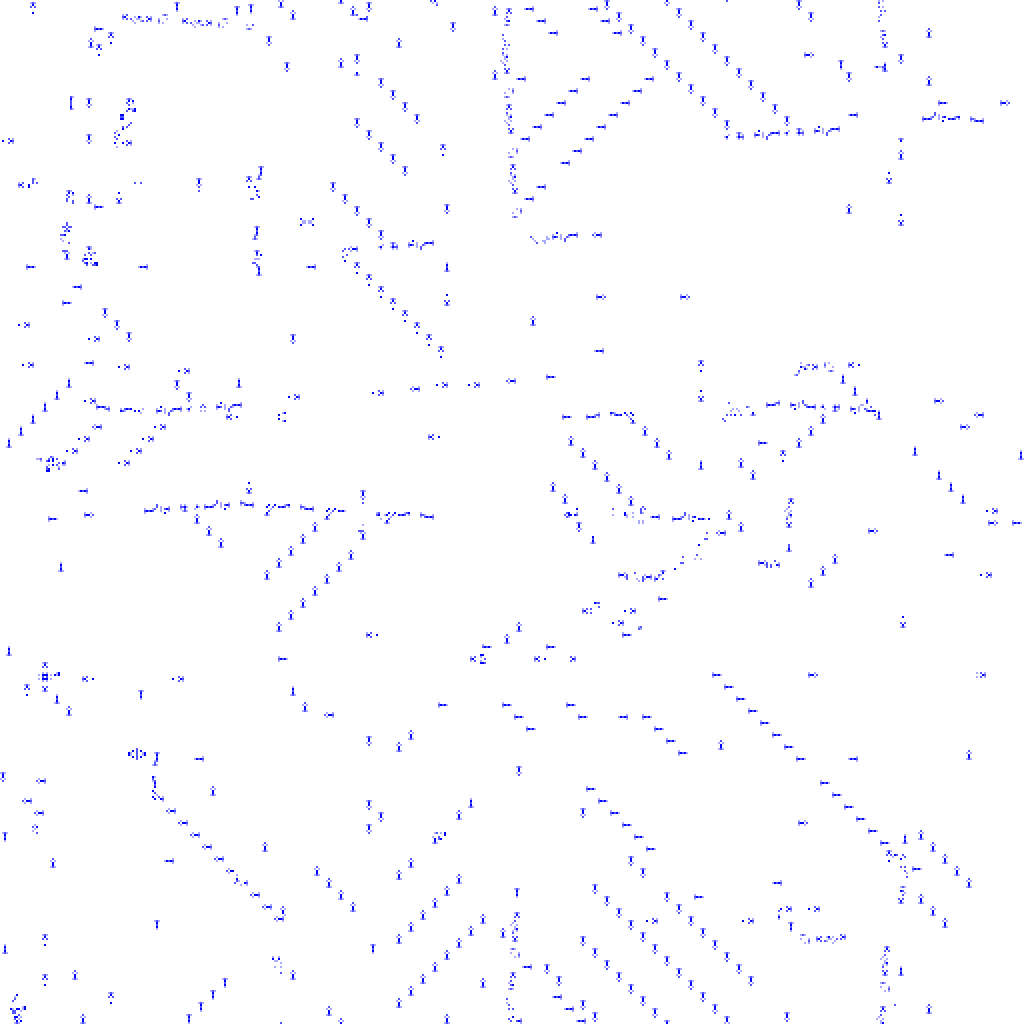}}
  \caption{\label{subfig:mult_glider_larger} $512 \times 512$ cells}
  \end{subfigure}
  \begin{subfigure}{.48\linewidth}
    \fbox{\centering
      \includegraphics[width=\linewidth]{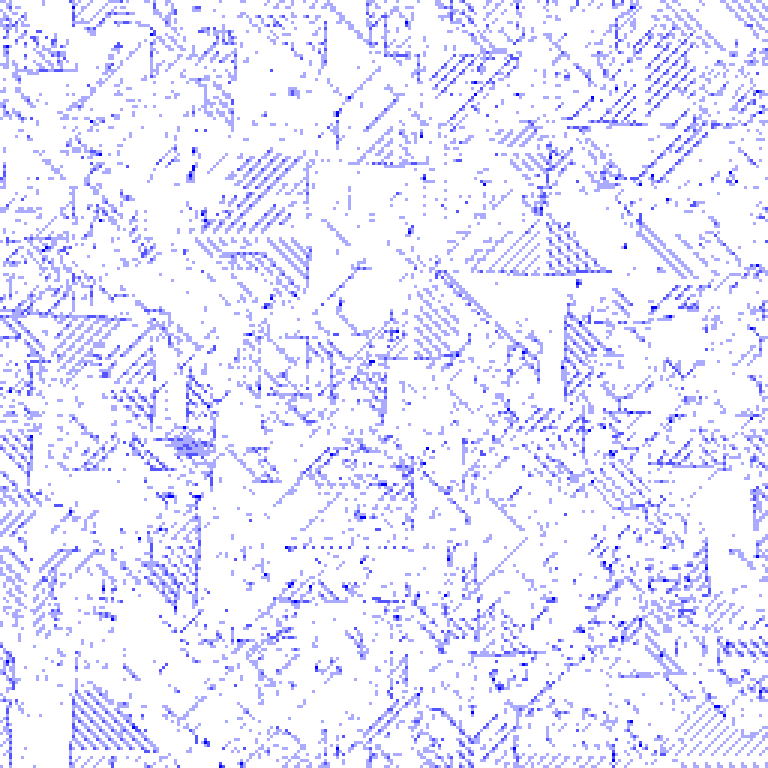}}
    \caption{\label{subfig:waves_b} $2048 \times 2048$ cells coarse-grained to $256 \times 256$.}
  \end{subfigure}
  \hfill
  \begin{subfigure}{.48\linewidth}
    \fbox{\centering
      \includegraphics[width=\linewidth]{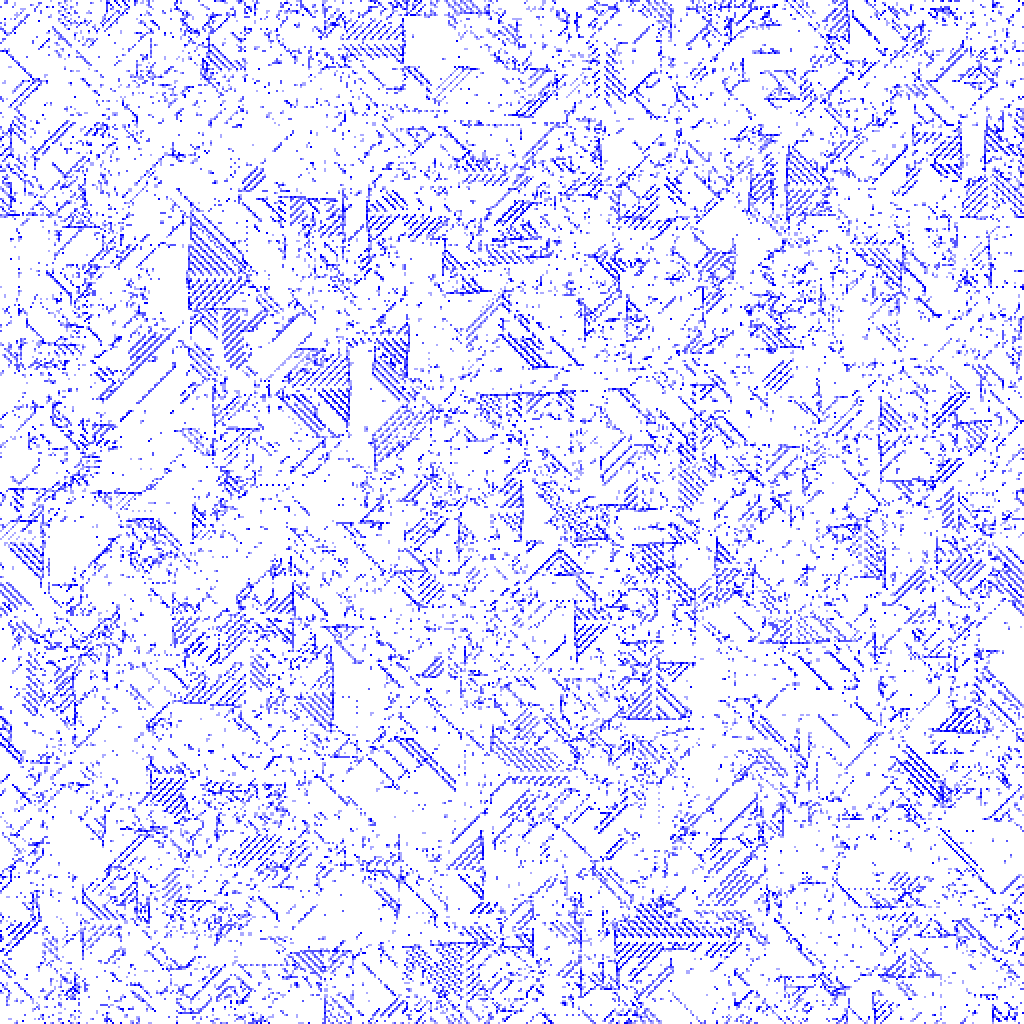}}
    \caption{\label{subfig:waves_c} $4096 \times 4096$ cells coarse-grained to $256 \times 256$.}
  \end{subfigure}
  \caption{\label{fig:dynamics_levels} \textbf{Changing CA dynamics at multiple
      scales}. (a) shows a single glider, oscillating between 3 positions. Such
    gliders emerge spontaneously from a random initialization of a small grid as
    shown in (b). When scaling the grid up, trails of gliders begin to appear,
    creating moving straight and diagonal lines as shown in (c). Scaling-up even
    more, individual gliders are not visible anymore, as shown in (d). In an
    even larger grid, shown in (e), many more triangular-shaped waves travel and
    collide with each other. Please note that (d) and (e) are coarse-grained to
    $256 \times 256$, otherwise the patterns are not visible.}
\end{figure}

Figure~\ref{fig:qualitative} is a qualitative comparison of coarse-graining
methods. This cellular automaton was selected from the experimental dataset.
When simulated on large grids, it generates large linear structures that are
4-cells wide. These structure disappear after downscaling by averaging because
the background dominates the average. Other methods correctly highlight these
structures when downscaling the grids by a factor of 8. In
Figure~\ref{fig:dynamics_levels}, we show another rule that was selected for its
high complexity score at multiple coarse-graining scales from our dataset. The
CA has significantly different dynamics depending on the chosen scale.
Example~\ref{subfig:single} is a spontaneously occurring stable oscillating
glider with period 3. Large structures emerge from these simple gliders when
observing large grids. The online project page\footnote{\projecturl} shows
animated example for Figure~\ref{fig:dynamics_levels}, emphasizing the advantage
of using coarse-graining for visualization.

A crucial advantage of the frequency histogram method is its speed and ease of
implementation compared to autoencoders. Other than a few hyper-parameters for
partitioning the histogram, no training or tuning is needed to produce the
coarse-grained output.

\section{Conclusion\label{sec:conclusion}}

We intend to use these coarse-graining methods to find cellular automata (CA) which
exhibit interesting behaviors at multiple scales.
Figure~\ref{fig:dynamics_levels} shows an example of such a CA\@. We observe
various dynamics depending on the scale, from simple oscillating gliders to
large wave-like patterns composed of thousands of gliders. It demonstrates that
observing multi-scale behaviors within those automata is possible. The existence
of 2D cellular automata with disordered behaviors at the smallest level but
organized at coarser scales, similar to hidden patterns in rule 18, would also
be of great interest.

Cellular automata are powerful computational models. Some of them have been
shown to be Turing-complete, and can thus be expected to support arbitrarily
complex computations~\citep{berlekampWinningWaysYour2001,
  cookUniversalityElementaryCellular2004}. Naturally, most interesting CAs
spontaneously generate a fraction of available computations at a time, usually
supporting a few stable oscillators or moving structures. Proofs of universality
for these CAs required careful design of computational devices out of these
stable oscillators and structures, resulting in very brittle and inefficient
universal computers. In practise, only elementary functions --- such as density
classification, binary addition, etc. --- can be implemented. This requires
searching for CA rules specifically targeted at a particular
function~\citep{mitchellEvolvingCellularAutomata1996, wolframNewKindScience2002,
  sapinResearchCellularAutomaton2003}. Hierarchies are central to naturally
occurring complex phenomena~\citep{simonArchitectureComplexity1962}, and may be
required for robust and complex processes to emerge in CAs.

Viewing space-time diagrams of cellular automata is akin to visualizing a
foreign computer design. Cellular automata are manipulating information,
registers and instructions in parallel in the form of cell states. We believe
visualization tools proposed in this paper can help understand computations in
those unconventional computers. By reducing available information to its
essential parts, we attempt to distill the content of the space-time diagram
with as little prior information as possible. Future work could focus on
identifying some known simple computational primitives within cellular automata
and understanding how our visualization can help to find them.

These methods also enable apprehending large grid sizes for which even image
processing algorithms begin to show limitations. Complexity metrics and CA
classification techniques can be extended to these reduced large grids and could
lead to the discovery of CAs with --- similar to life and physical processes ---
significantly different dynamics at multiple scales that could in turn be a
basis for artificial evolution.

{\footnotesize \paragraph{Acknowledgements}\label{sec:ack}
This work was partially supported by ERC grant LEAP No. 336845, CIFAR Learning
in Machines \& Brains program and the EU Structural and Investment Funds,
Operational Program Research, Development and Education under the project
IMPACT (reg. no. CZ.02.1.01/0.0/0.0/15003/0000468).}

\footnotesize
\bibliographystyle{apalike}
\bibliography{library}

\end{document}